\documentclass[aps,pre,preprint,superscriptaddress,longbibliography]{revtex4-2}

\usepackage{amsmath}
\usepackage{amssymb}
\usepackage{appendix}
\usepackage{bm}
\usepackage{color}

\usepackage{graphicx}
\usepackage[colorlinks=true, allcolors=blue]{hyperref}
\usepackage{subfigure}

\begin{document}

\title{Spontaneous symmetry-breaking in equilibrium tree-packing configurations of a kinetically constrained cubic-lattice system}

\author{Hai-Jun Zhou}
\affiliation{
  Institute of Theoretical Physics, Chinese Academy of Sciences, Zhong-Guan-Cun East Road 55, Beijing 100190, China
}
\affiliation{
  School of Physical Sciences, University of Chinese Academy of Sciences, Beijing 100049, China
}
\affiliation{
  Institute for Advanced Physical Studies, Zhejiang University, Hangzhou 310027, China}

\date{\today}

\begin{abstract}
We explore kinetic-constraint induced thermodynamic phase transition in the cubic lattice, employing the Fredrikson-Anderson spin model with hyperparameter $K=2$ as a representative kinetic system. Each lattice site may flip its binary occupation state if at most one of its six nearest neighbors is currently occupied. The whole set of microscopic configurations that are kinetically connected with the fully empty one is described by an equilibrium partition function with a single global constraint, that is, the occupied sites do not form closed loops but instead organize into different tree components in the lattice.  We discover a continuous thermodynamic gas--crystal phase transition in the cubic system and determine the critical chemical potential $\mu^* \approx -3.252$, at which the occupied sites of the equilibrium tree-packing configurations start to prefer one of the two nested cubic sublattices. This thermodynamic phase transition is absent in the two-dimensional square lattice.    
\end{abstract}

\maketitle

\subsection*{Introduction}

Kinetically constrained spin models are discrete-state lattice toy systems for understanding the glass dynamics. The kinetic constraints of such models can be direction-symmetric such as in the Fredrickson-Andersen (FA) kinetic model~\cite{Fredrickson-Andersen-1984} or direction-asymmetric such as in the East model~\cite{Jackle-Eisenger-1991}. In the FA model, for example, each lattice site has a binary state (empty or being occupied) which is free to flip only if it has less than $K$ occupied nearest neighbors (otherwise the site is temporarily blocked in its present state). This simple model and the other kinetic lattice systems exhibit very rich dynamical behaviors resembling those of real-world glass-forming liquids~\cite{Ritort-Sollich-2003,Faggionato-etal-2013,Perrupato-Rizzo-2025b}.

The equilibrium thermodynamic properties of kinetically constrained spin models, however, have rarely been discussed~\cite{Foini-etal-2012,Perrupato-Rizzo-2023,Zhou-2024}.  For the FA model defined on $D$-dimensional hypercubic lattices, it was rigorously proven that, if the kinetic threshold $K \geq (D+1)$, the whole space of microscopic configurations is ergodic in the kinetic sense, except for a vanishingly small fraction of special configurations~\cite{vanEnter-1987,Schonmann-1990}. The thermodynamics of an infinite system is then identical to that of a non-interacting lattice gas, and the diverging relaxation time at high occupation densities must purely be a kinetic effect. At the other extreme of $K=1$, the kinetic constraint is so severe that the FA model is equivalent to a lattice gas with local hard-core repulsive interactions, and it has a continuous phase transition between the gas phase and the crystalline phase in two- and three-dimensions~\cite{Baxter-1980}. A continuous phase transition between the ordered crystalline phase and a disordered glass phase was also recently confirmed by computer simulations in the ground-state configuration subspace of the cubic lattice~\cite{Liu-Zhou-etal-2026}. For the $K=1$ FA model defined on random graphs, the system may enter into the spin glass phase as the density of occupied sites becomes sufficiently high~\cite{Zhang-Zeng-Zhou-2009}.

In the present paper we investigate the thermodynamic properties of the FA kinetic model with control parameter $K=2$ on cubic lattices and two-dimensional square lattices. The $K=2$ kinetic rule brings strongly non-local and non-reciprocal effective interactions between the lattice sites. Starting from an initial configuration with all the sites being empty, the kinetic constraint of blocking any site with two or more occupied nearest neighbors from flipping state leads to the emergence of a global (topological) property: the occupied sites will never form a closed loop in the lattice~\cite{Zhou-2024}. The occupied sites organize into many mutually disconnected tree components, and the empty sites form a feedback vertex set. The $K=2$ FA model can therefore be investigated as a feedback vertex set problem and it is known to have spin glass phase transitions in random graphs~\cite{Qin-etal-2016}. Because the absence of occupied loops is a global structural property, this problem in finite dimensions is very difficult to tackle by analytical means.

Here we demonstrate by extensive numerical simulations that it has a continuous gas--crystal equilibrium phase transition in the cubic lattice. Our work reveals a foundamental thermodynamic property of the three-dimensional FA kinetic system. The symmetry between the two nested cubic sublattices breaks down spontaneously as the chemical potential drops below the critical value $\mu^* \approx -3.252$ and the occupied sites of the two sublattices start to play different roles in the tree components. In contrast, we find that no thermodynamic phase transition occurs in the square lattice and the system is always in the disordered and homogeneous gas phase. An open issue for the near future is to investigate whether this thermodynamic difference between three-dimension and two-dimension is associated with qualitative differences in the dynamical behavior.

\subsection*{Model and numerical approach}

We consider cubic lattices with periodic boundary conditions along all the three dimensions. The total number of sites is $N = L^3$ with $L$ being the lattice side length and each site $i \in \{1, \ldots, N\}$ is connected to six nearest neighbors.  We set $L$ to be even so that the whole system contains two nested cubic sublattices ($A$ and $B$), preserving a basic structural property of real-world cubic lattices. All the nearest neighbors of a site of one sublattice belong to the other sublattice. Each lattice site $i$ has a binary state $c_i = 0$ (being empty) or $c_i = 1$ (being occupied). Under the FA kinetic rule, a site $i$ is free to flip state at any given moment if (and only if) it has less than $K$ occupied nearest neighbors. We fix $K=2$ in the present work (the simpler $K=1$ case was studied in Ref.~\cite{Liu-Zhou-etal-2026}).

Let us denote by $\mathcal{C}_1$ the whole set of all the microscopic configurations $\vec{\bm{c}} = (c_1, \ldots, c_N)$ that are reachable from the initial completely empty configuration $(0,\ldots, 0)$ through a kinetically allowed single-site-flipping trajectory. We call $\mathcal{C}_1$ the fully unfrozen kinetic cluster~\cite{Zhou-2024}. It is obvious that the occupied sites in any configuration $\vec{\bm{c}}$ of $\mathcal{C}_1$ must \emph{not} form closed loops because otherwise all the sites in these loops will be permanently occupied and $\vec{\bm{c}}$ will be kinetically disconnected from $(0,\ldots, 0)$. It is also easy to verify that, if the occupied sites of a microscopic configuration $\vec{\bm{c}}$ do not form closed loops, then $\vec{\bm{c}}$ is kinetically connected with $(0,\ldots, 0)$ and therefore belongs to $\mathcal{C}_1$~\cite{Zhou-2024,Perrupato-Rizzo-2023}. The thermodynamic properties of the fully unfrozen kinetic cluster can then be investigated through the following partitition function
\begin{equation}
  Z( \mu ) = \sum\limits_{ \vec{\bm{c}} \in  \{0, 1\}^N} I\bigl( \vec{\bm{c}}\;\; \textrm{has no occupied loop} \bigr) \prod\limits_{i=1}^{N} e^{  - \mu c_i } \; ,
  \label{eq:Zmu}
\end{equation}
which is a summation over all the $2^N$ possible microscopic configurations. Here $\mu$ is the chemical potential controlling the density of occupied sites $\rho$ ($= \sum_i c_i / N$); the indicator function $I = 1$ if configuration $\vec{\bm{c}}$ has no occupied loop and $I=0$ if otherwise.

Equation (\ref{eq:Zmu}) is equivalent to a weighted sum over all the tree-packing configurations, the occupied sites of which form many mutually separated tree components (free of any closed loops)~\cite{Qin-etal-2016}. Let us emphasize that that no-occupied-loop is a global (topological) property of the microscopic configurations and it does not locally restrict the number of occupied nearest neighbors of individual sites, in contract to lattice glass models~\cite{Rivoire-etal-2004} and the $K=1$ (hard-core) kinetic systems~\cite{Baxter-1980,Liu-Zhou-etal-2026}.

We use chemical potential ($\mu < 0$) as the control parameter and run an efficient Markov-Chain Monte Carlo dynamics to sample a large set of equilibrium configurations. Besides single-site flips, we also allow state-swapping between a pair of sites~\cite{Fan-Zhou-2023}, and we maintain a set $\Gamma_0$ of flippable empty sites. Given any configuration $\vec{\bm{c}}$, if a site $i$ is empty and if flipping it to $c_i=1$ will \emph{not} lead to the formation of a closed loop of occupied sites in the lattice, then $i$ is a flippable empty site ($i \in \Gamma_0$). We assign each occupied site $j$ a tree index to distinguish the different tree components of occupied sites. If an empty site $i$ is connected to two or more occupied sites with the same tree index, then it is temporarily not flippable ($i\notin \Gamma_0$). The tree indices of all the occupied sites are updated after each modification of the configuration $\vec{\bm{c}}$, and we have made efforts to make this process as quick as possible.

At each elementary step of the stochastic dynamics, we first pick a site $i$ uniformly at random from the lattice and check whether it is currently occupied  or empty. (1) If $c_i=1$, we flip its state with certainty ($c_i: 1\rightarrow 0$) and then add site $i$ and maybe some additional empty sites to the flippable empty set $\Gamma_0$; and with probability $p_{\textrm{swap}}= (1 - e^{\mu})$ we complement this flip by flipping an empty site $j$ ($c_j: 0 \rightarrow 1$), with $j$ chosen uniformly at random from the updated set $\Gamma_0\backslash\{i\}$ (if $\Gamma_0$ contains only site $i$, we take $j=i$), and then update $\Gamma_0$ once again. (2) If $c_i=0$ and $i \in \Gamma_0$ we just flip this site with certainty ($c_i: 0 \rightarrow 1$) and then update the set $\Gamma_0$; if $c_i=0$ but $i \notin \Gamma_0$, we let site $i$ to remain in the empty state. The evolution time is increased by a tiny amount $1/N$ after each elementary step of the stochastic dynamics (one unit time corresponds to $N$ consecutive elementary steps of updating trials). 

This above stochastic dynamics obeys detailed balance. We have checked that the autocorrelation time of the magnetic order parameter $m$ (see blow) is less than $10^4$ time units for all the $\mu$ values considered in this work. We run $64$--$65$ independet stochastic evolution trajectories at each fixed $(L, \mu)$ point. After the stochastic dynamics has reached equilibrium, we sample at least $3.2\times 10^6$ equilibrium tree-packing configurations (at unit time interval) from these independent runs, and then carry out conventional statistical analysis on them~\cite{Amit-MartinMayor-2005}. 

\subsection*{Spontaneous symmetry-breaking}

\begin{figure}
  \centering
  \subfigure[]{
    \includegraphics[angle=270,width=0.45\linewidth]{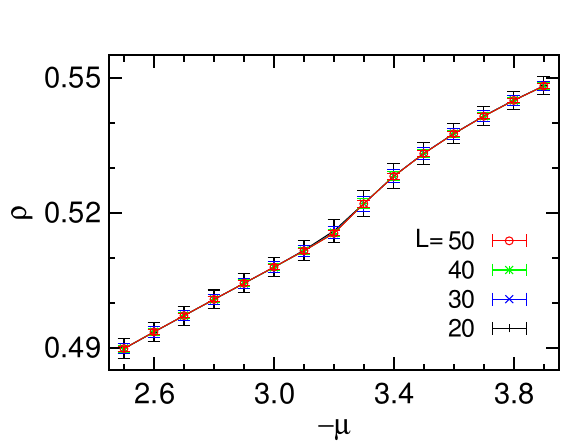}
    \label{fig:Cubic:Rho}
  }
  \subfigure[]{
    \includegraphics[angle=270,width=0.45\linewidth]{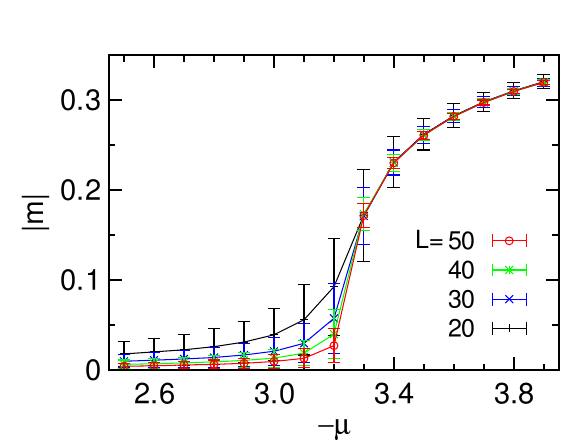}
    \label{fig:Cubic:Mabs}
  } \\
  \subfigure[]{
    \includegraphics[angle=270,width=0.45\linewidth]{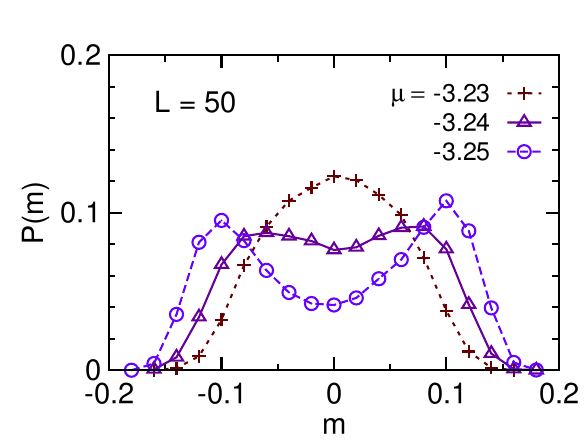}
    \label{fig:Cubic:Pm}
  }
  \subfigure[]{
      \includegraphics[angle=270,width=0.45\linewidth]{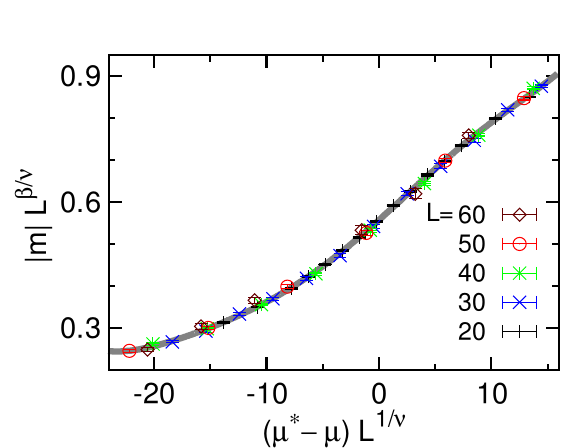}
      \label{fig:Cubic:FSS}
  }
  \caption{
  Numerical results obtained on periodic cubic lattices of various side lengths $L$ at different chemical potential values $\mu$. (a) Mean value of occupation density $\rho$ and its standard deviation. (b) Mean value of absolute magnetization $|m|$ and its standard deviation. (c) Probability profiles $P(m)$ of magnetization $m$ for the system with side length $L=50$. (d) Confirmation of the finite-size-scaling relation (\ref{eq:fss}) between $|m| L^{\beta/ \nu}$ and $L^{1/\nu}(\mu - \mu^*)$; errorbars mark the rescaled standard errors of the mean, and the gray thick line is the scaling function $f(x)$; the critical chemical potential is $\mu^* \approx -3.252$.
  }
  \label{fig:Cubic}
\end{figure}

Figure~\ref{fig:Cubic:Rho} shows how the mean occupation density $\rho$ of equilibrium tree-packing configurations changes with the chemical potential $\mu$. We clearly see that the trend of $\rho(\mu)$ changes at $\mu \approx -3.25$. After examining some individual equilibrium configurations, we realize that this apparent singularity is caused at the microscopic level by a spontaneous symmetry-breaking between the two nested cubic sublattices $A$ and $B$. Let us define a magnetization order parameter $m$ as
\begin{equation}
 m = \frac{1}{N} \Bigl[ \sum\limits_{i \ \in \  \textrm{sublattice A}} c_i 
 - \sum\limits_{j\ \in \ \textrm{sublattice B}} c_j \Bigr] \; .
 \label{eq:m}
\end{equation}
If the occupied sites are equally distributed in the two sublattices then $m$ will be of order $N^{-1/2}$; if one sublattice is much more densely occupied than the other one, then $m$ will be of order unity.

Figure~\ref{fig:Cubic:Mabs} shows how the absolute magnetization $|m|$ changes with $\mu$. The numerical results clearly indicate that $|m|$ will decay to zero at $L \rightarrow \infty$ for $\mu \geq -3.2$ but it will approach a positive constant for $\mu \leq -3.3$. A thermodynamic phase transition is expected to occur at certain critical chemical potential $\mu^* \in (-3.3, -3.2)$. The existence of such a phase transition is further confirmed by the shape change of the probability profile $P(m)$ of the magnetization values. For example, for the finite system of side length $L=50$, we find $P(m)$ has only a single peak at $m \approx 0$ at $\mu = -3.23$; when $\mu$ decreases to $-3.24$ this single peak splits into two symmetric peaks at $m \approx \pm 0.06$ and itself changes to be a shallow valley; as $\mu$ further decreases to $-3.25$ the valley at $m=0$ become much deeper and the two symmetric peaks become further separated to $m \approx \pm 0.10$ [Fig.~\ref{fig:Cubic:Pm}]. This evolution behavior of $P(m)$ suggests that the phase transition is a continuous one, simular to the emergence of magnetism in the three-dimensional Ising model.

We perform the conventional finite-size-scaling (FSS) analysis to estimate the critical point $\mu^*$ for an infinitely large system~\cite{Amit-MartinMayor-2005}. In the vicinity of $\mu^*$ we assume that the mean value of the absolute magentization $|m|$ is governed by the scaling law
\begin{equation}
    | m | \, = \, L^{-\beta / \nu} f\bigl( L^{1/\nu} (\mu - \mu^*) \bigr) \; ,
    \label{eq:fss}
\end{equation}
with $\beta$ and $\nu$ being two critical exponents. The scaling function $f(x) = \sum_{n=0}^q a_n x^n$ is a $q^{\textrm{th}}$-order polynomial with fitting parameters $a_n$ (the optimal $q=6$). We obtain $47$ mean values of $|m|$ at several chemical potentials $\mu \in (-3.34, -3.16)$ for various systems of size $L \in \{20, 30, 40, 50, 60\}$ by extensive numerical simulations. By minimizing the rescaled chi-square loss function as in Ref.~\cite{Liu-Zhou-etal-2026}, we estimate the critical point to be $\mu^* = -3.25160(8)$ and the critical exponents to be $\beta = 0.2864(1)$ and $\nu = 0.5969(2)$ [Fig.~\ref{fig:Cubic:FSS}]. We notice that the estimated critical exponents appear to be distinct from those of the three-dimensional Ising model (which are $\beta \approx 0.326$ and $\nu \approx 0.630$~\cite{Jaster-etal-1999,Chang-etal-2025}). The revealed continuous phase transition of the $K=2$ kinetic system may belong to a different universality class, but more efforts are definitely needed to clarify this universality issue.

\begin{figure}
  \centering
  \subfigure[]{
    \includegraphics[angle=270,width=0.45\linewidth]{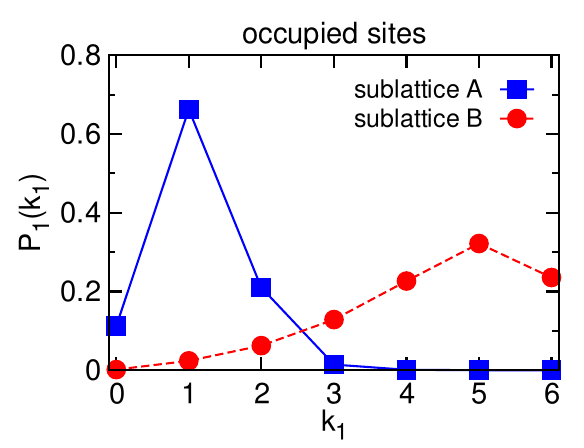}
    \label{fig:CDL50:Occupy}
  }
  \subfigure[]{
    \includegraphics[angle=270,width=0.45\linewidth]{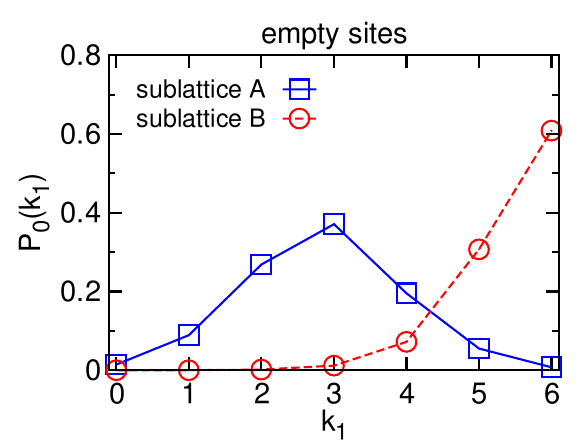}
    \label{fig:CDL50:Empty}
  }
  \caption{
  Probabilities $P_1(k_1)$ of an occupied site (a) and $P_0(k_1)$ of an empty site (b) to have $k_1$ occupied nearest neighbors in an equilibrium tree-packing configuration. The periodic cubic lattice has side length $L=50$, and the chemical potential is fixed to $\mu = -4.0$.  Results obtained on the sites of the dense sublattice $A$ and the sparse sublattice $B$ are distinguished by squares and circles. Filled (open) symbols correspond to occupied (empty) sites.
  }
  \label{fig:CDL50}
  \end{figure}

In the crystalline phase, the occupied sites of sublattice A and sublattice B take different structural roles in the tree components of occupied sites. For example, at chemical potential $\mu = -4.0$, if an occupied site belongs to the more densely occupied sublattice (say $A$), it is most likely to be a leaf site (having only one occupied nearest neighbor), while an occupied site of the more sparsely occupied sublattice $B$ is most likely to be a junction point of $4$--$6$ tree branches [Fig.~\ref{fig:CDL50:Occupy}]. An empty site of the dense sublattice $A$ most likely has $2$--$4$ occupied nearest neighbors, while an empty site of the sparse sublattice $B$ is most likely to have all its six nearest neighbors being occupied [Fig.~\ref{fig:CDL50:Empty}]. These distinctions explain why flipping an empty site of sublattice $B$ is more difficult than flipping an empty site of sublattice $A$.

\subsection*{No phase transition in square lattice}

We also carry out the same thermodynamic study on square lattices of various even side lengths $L$. Such two-dimensional periodic lattices also have two nested square sublattices but very surprisingly, we find that there is no spontaneous symmetry-breaking between these two sublattices (see supplementary information). The magnetic order parameter $m$ is distributed around zero at all chemical potential values $\mu$; the mean density $\rho$ of occupied sites approaches the maximum value $2/3$ smoothly as $\mu$ decreases; and all the lattice sites are equally likely to be occupied, characteristic of a disordered gas phase. Therefore the minimum spatial dimension appears to be three to have a thermodynamic phase transition in the $K=2$ FA kinetic system.

\subsection*{Discussion}

In summary, we discovered a continuous thermodynamic phase transition in the Fredrickson-Andersen kinetically constrained spin model with control parameter $K=2$.  The equilibrium tree-packing configurations of the three-dimensional cubic lattice start to be partially ordered at the critical chemical potential $\mu^* \approx -3.252$, while this phase transition is absent in the two-dimensional square lattice. Further work is needed to confirm that this three-dimensional gas--crystal phase transition does \emph{not} belong to the universality class of the ferromagnetic Ising model.

In the near future it will be very interesting to run single-site-flipping dynamics under the FA kinetic rule on cubic lattices to determine whether this equilibrium gas--crystal phase transition is kinetically accessible; if it turns out to be kinetically avoided at sufficiently negative chemical potentials, the system must be out of equilibrium. Since the square and cubic lattice systems have qualitatively distinct thermodynamic properties, we expect that they will have distinct dynamical properties under the FA kinetic rule, and hopefully our work may stimulate serious comparative studies along this direction. It is also necessary to extend the present work to other lattice structures and to higher dimensions. We may also consider random initial configurations with a positive fraction of occupied sites as in Ref.~\cite{Liu-Zhou-etal-2026} and investigate the emergence of a thermodynamic glass phase.

Extending the present work to the cases with $K \geq 3$ is another interesting future direction. The equilibrium statistical mechanics problem is equivalent to the $K$-core attack problem~\cite{Zhou-2024,Guggiola-Semerjian-2015,Zhou-2022}, which is much more challenging to perform computer simulations.

\vskip 0.2cm

\subsection*{Acknowledgments}

The author thanks Yejia Chen and Ruifeng Liu for helpful discussions on finite-size-scaling analysis. The following funding supports are acknowledged: National Natural Science Foundation of China Grants No.~12247104, No.~T2541021  and No.~12447101. Numerical simulations were carried out at the HPC cluster of ITP-CAS and the BSCC-A2 platform of the National Supercomputer Center in Beijing.


%

\clearpage


\begin{appendix}

  \begin{center}
 \large{Supplemantary Information}
  \end{center}

  \vskip 1.0cm
  
\subsection*{Data availability}
    
The data that support the findings of this article will be made openly available at Science Data Bank (https://doi.org/10.57760/sciencedb.45778).

\subsection*{Finite-size-scaling analysis}
\label{app:fss}

To perform finite-size-scaling (FSS) analysis, we repeat our equilibrium Monte Carlo stochastic process a total number of $S$ times at various fixed lattice side lengths $L_s \in \{20, 30, 40, 50, 60\}$ and chemical potentials $\mu_s$ at the vicinity of $-3.25$. These numerical experiments are indexed by sample index $s \in \{1, 2, \ldots, S\}$ with sample size $S = 47$. For each numerical experiment at $(L_s, \mu_s)$, we sample $3.2 \times 10^6 - 1.0 \times 10^7$ equilibrium tree-packing microscopic configurations at unit time interval. We divide the data sequence into $64$--$65$ non-overlapping blocks and obtain the empirical mean  of the absolute magnetization $|m|$ for each data block. We then use these block mean values to compute the mean absolute magnetization $\overline{|m|}_s$ and the standard error $\sigma_s$ of the sample at $(L_s, \mu_s)$.

If the true mean value of the absolution magnetization $|m|$ follows the FSS relationship (\ref{eq:fss}), then the joint probability distribution of observing the set of $S$ empirical data points $\{\overline{|m|}_s\}$ is
\begin{equation}
\begin{aligned}
  P\bigl(\{\overline{|m|}_s\} \bigr) \, \propto \,
  & \prod\limits_{s=1}^S \exp\biggl[ - \frac{ \big( \overline{|m|}_s - L_s^{-\beta/\nu} \sum_{n=0}^{q} a_n [L_s^{1/\nu} (\mu - \mu^*)]^n  \bigr)^2}{2 \sigma_s^2} \biggr] 
  \\
  = \, & \exp\Bigl[ - \mathcal{L}\bigl(\mu^*, \beta, \nu, \{a_n\}\bigr) \Bigr] \; .
\end{aligned}
\label{eq:Pdata}
\end{equation}
Our goal is to determine the values of $\mu^*$, $\beta$, $\nu$ and the $(q+1)$ polynomial coefficients $a_n$ by minimizing the minus logarithmic probability $\mathcal{L}$,
\begin{equation}
\mathcal{L}\bigl( \mu^*, \beta, \nu, \{a_n\}\bigr) \, = \, \sum\limits_{s=1}^{S} \frac{ \bigl( \overline{|m|}_s - L_s^{-\beta/\nu} \sum_{n=0}^{q} a_n [ L_s^{1/\nu} (\mu_s - \mu^*) ]^n \bigr)^2}{ 2 \sigma_s^2} \; .
\label{eq:L}
\end{equation}
We refer to $\mathcal{L}$ as the loss function. 

The global minimum point of $\mathcal{L}$ can be easily reached by a simple stochastic annealing process in the ($q+4$)-dimensional space of fitting parameters. The optimal value of $q$ is determined repeatedly running this annealing algorithm at different values of $q$ to get the one with the minimum value of the rescaled chi-square $\chi^2$, which is defined as
\begin{equation}
  \chi^2 \, = \, \frac{2 \mathcal{L}}{S - (q+4)}
  \, = \, \frac{1}{S - q - 4} \sum\limits_{s=1}^{S} \frac{ \bigl( \overline{|m|}_s - L_s^{-\beta/\nu} \sum_{n=0}^{q} a_n [ L_s^{1/\nu} (\mu_s - \mu^*) ]^n \bigr)^2}{\sigma_s^2} \; .
  \label{eq:chisq}
\end{equation}
In the above expression, $S - (q+4)$ is the degrees of freedom (number of data samples minus the total number of fitting parameters). For our FSS analysis with $S = 47$ sample points, our FSS analysis achieves a minimum value of $\chi^2 \approx 1.97$ at $q = 6$. 

To estimate the uncertainty of the inferred critical chemical potential $\mu^*$, we compute the second-order partial derivative of $\mathcal{L}$ with respect to $\mu^*$ at the global minimum point according to
\begin{equation}
\frac{\partial^2 \mathcal{L}}{\partial (\mu^*)^2} \, =  \, 
\sum\limits_{s=1}^S \frac{L_s^{2/\nu}}{\sigma_s^2} \Bigl[ 
\bigl( L_s^{-\beta/\nu} \sum_{n=0}^q n a_n [L_s^{1/\nu}( \mu_s - \mu^*) ]^{n-1} \bigr)^2 
- L_s^{-\beta/\nu} \Delta_s \sum\limits_{n=0}^q n (n - 1 ) a_n [ L_s^{1/\nu}(\mu_s - \mu^*) ]^{n-2} \Bigr] \; ,
\end{equation}
where $\Delta_s$ is a short-hand notation for the difference between the empirical mean and the prediction, 
\begin{equation}
\Delta_s \, = \, \overline{|m|}_s - L_s^{-\beta / \nu} 
\sum\limits_{n=0}^q a_n \bigl[ L_s^{1/\nu} (\mu_s - \mu^*) \bigr]^n \; .
\end{equation}  
By assuming a uniform prior distribution for the critical chemical potential $\mu^*$, we know from Eq.~(\ref{eq:Pdata}) that the posterior probability of $\mu^*$ given the $S$ empirical data points is $Q( \mu^*) \propto \exp( - \mathcal{L} \bigr)$. By expanding $\mathcal{L}$ as a quadratic function of $\mu^*$ at the vicinity of the global minimum point, we obtain the standard error on $\mu^*$ as $1/ \sqrt{\partial^2 \mathcal{L} / \partial (\mu^*)^2}$. 

Similarly, we can estimate the uncertainty of the inferred critical exponent $\beta$ and that of $\nu$ by computing the following two second-order derivatives:
\begin{equation}
\frac{\partial^2 \mathcal{L}}{\partial \beta^2} \, =  \, \frac{1}{\nu^2}
\sum\limits_{s=1}^M \frac{(\ln L_s)^2}{\sigma_s^2} \Bigl[ 
\bigl( L_s^{-\beta/\nu} \sum_n a_n [L_s^{1/\nu}( \mu_s - \mu^*) ]^{n} \bigr)^2 
- L_s^{-\beta/\nu} \Delta_s \sum\limits_{n} a_n [ L_s^{1/\nu}(\mu_s - \mu^*) ]^{n} \Bigr] \; ,
\end{equation}
and
\begin{equation}
\begin{aligned}
\frac{\partial^2 \mathcal{L}}{\partial \nu^2} \, = & \, 
\frac{(\ln L_s)^2}{\nu^2} \sum\limits_{s=1}^M \frac{1}{\sigma_s^2} \Bigl[ 
\bigl( L_s^{-\beta/\nu}  \sum_n  a_n (-\beta/\nu + n/\nu) [L_s^{1/\nu}( \mu_s - \mu^*) ]^{n} \bigr)^2 \\
& \quad  \quad \quad \quad
- (2 / \ln L_s) L_s^{-\beta/\nu}  \Delta_s \sum\limits_{n} a_n ( -\beta/\nu + n/\nu ) [ L_s^{1/\nu}(\mu_s - \mu^*) ]^{n} 
\\
& \quad \quad \quad \quad-  L_s^{-\beta/\nu} \Delta_s \sum_n a_n (-\beta/\nu + n/\nu)^2  [ L_s^{1/\nu} (\mu_s - \mu^*) ]^n
\Bigr] \; .
\end{aligned}
\end{equation}

\subsection*{Results for periodic square lattices}
\label{app:2Dsystem}

\begin{figure}[!h]
  \centering
  \subfigure[]{
    \includegraphics[angle=270,width=0.3\linewidth]{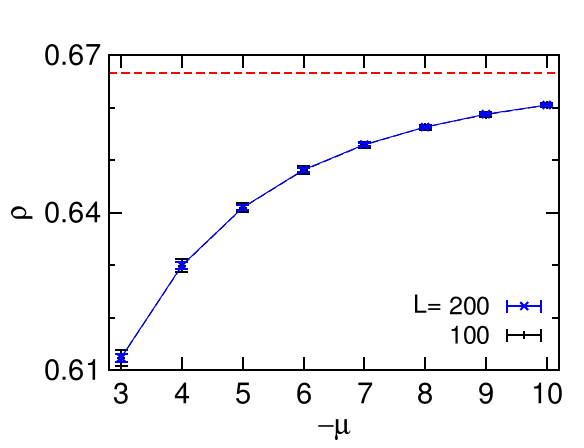}
    \label{fig:Square:rho}
  }
  \subfigure[]{
    \includegraphics[angle=270,width=0.3\linewidth]{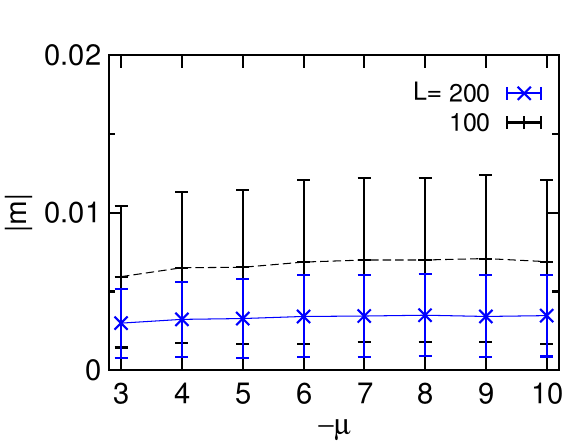}
    \label{fig:Square:absm}
  }
    \subfigure[]{
      \includegraphics[angle=270,width=0.3\linewidth]{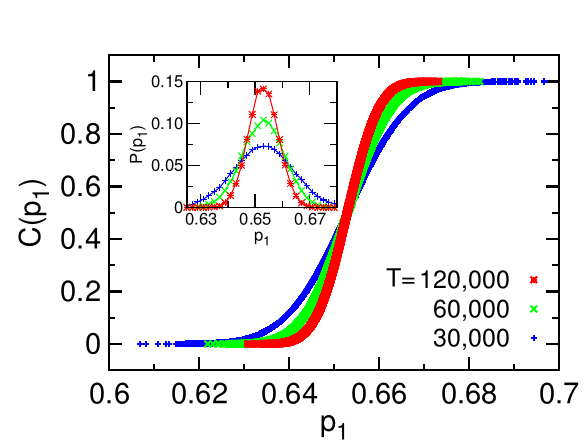}
          \label{fig:Square:p1}
  }
    \caption{
      Monte Carlo simulation results for the two-dimensional periodic lattices of side lengths $L=100$ and $L=200$. (a) Mean density $\rho$ of occupied sites versue chemical potential $\mu$; the dashed horizontal line marks the upper-bound density $2/3$. (b) Mean value of absolute magnetization $|m|$ and its standard deviation versue $\mu$. (c) Cumulative distribution $C(p_1)$ and probability profile $P(p_1)$ (inset) of $p_1$ averaged over an equilibrium trajectory of $T$ time units ($T=3\times 10^4, 6\times 10^4, 12 \times 10^4$) for the system of $L=200$; here $p_1$ is the fraction of times a randomly chosen site is observed to be occupied in these $T$ sampled configurations.
    }
  \label{fig:Square}
\end{figure}

We apply our equilibrium Monte Carlo algorithm to various periodic square lattices as well. Some typical results are shown in Fig.~\ref{fig:Square}. The mean density $\rho$ of occupied sites is a smooth function of chemical potential $\mu$ and it approaches the maximum value of $2/3$ as $\mu \rightarrow -\infty$ [Fig.~\ref{fig:Square:rho}]. The mean value of the absolute magnetization $|m|$ is produced by random fluctuations of $m$ arround zero at any value of $\mu$, and it decreases with the lattice side length $L$ [Fig.~\ref{fig:Square:absm}]. These results suggest that the equilibrium tree-packing configurations always belong to the disordered gas phase at any chemical potential $\mu$. There is no spontaneous symmetry-breaking and non equilibrium phase transition in this two-dimensional system.

To further confirm the absense of order, we compute the fraction $p_1(i)$ of times within an equilibrium trajectory of $T$ unit times for each lattice site $i$. The cummative distribution $C(p_1)$ of this fraction $p_1$ among all the $N=L^2$ sites of the lattice and the corresponding probability profile $P(p_1)$ are shown in Fig.~\ref{fig:Square:p1}, for trajectories sampled at very negetive chemical potential $\mu = -10$. As trajectory length $T$ increases, the $p_1$ values of different sites become more and more similar to each other and they concentrate around the mean occupation density $\rho \approx 0.65$ of the system. These results clear demonstrate that the system is homogeneous even when the sampled equilibrium configurations are close to the most densely occupied configurations.

\end{appendix}

\clearpage


\end{document}